\documentclass[b5paper,twoside]{jpconf}
\usepackage{graphicx}
\usepackage{epstopdf}

\begin{document}
\title[Progenitors of Nova Explosions]{Investigation of the Progenitors
\\ of Nova Explosions}

\author[F. Surina]{F. Surina, M. F. Bode, and M. J. Darnley}

\address{Astrophysics Research Institute, Liverpool John Moores University,

Twelve Quays House, Egerton Wharf, Birkenhead, CH41 1LD, UK}

\ead{mfs@astro.livjm.ac.uk}

\begin{abstract}
Recurrent novae (RNe) play an important role as one of the suspected progenitor systems of Type Ia supernovae (SNe) which are used as primary distance indicators in cosmology.  Thus, it is important to investigate the nature of their central binary systems to determine the relation between the parameters of the central system and outburst type, and finally ascertain the population of novae that might be available to give rise to the progenitors of Type Ia SNe. The details of the preliminary investigation looking for characteristics that may distinguish recurrent and classical novae (CNe) systems, the selection of initial targets for detailed study, and preliminary results are presented. We identified 10 suspected RNe among the Galactic CNe for investigation with the Liverpool Telescope population and our initial photometric observations of their quiescent systems suggest 2 may belong to the RS Oph type and 2 to the U Sco type RNe. Spectroscopic follow-up is now underway.
\end{abstract}

\section{Introduction}
Classical novae (CNe) are interacting binary systems whose outbursts are powered by a thermonuclear runaway on the surface of a white dwarf. Meanwhile secondary stars are filling their Roche lobe and transferring material onto the white dwarf primary stars via accretion  disks [1]. Recurrent novae (RNe) are, by definition, CNe with multiple recorded outbursts and may contain evolved secondaries [2].

There are 10 known Galactic RNe and they group themselves into three distinct subtypes according to ref.[2]:
\begin{enumerate}
\item {\itshape RS Oph type (T CrB, RS Oph, V745 Sco and V3890 Sgr)\/} have long orbital periods ($\sim$$10^{2}$ d). Their red giant secondary stars make them similar to symbiotic systems. The mass transfer rate and outburst ejection velocities are high. Light curves decline rapidly after outbursts. This sub-type of RNe has strong evidence for interaction between ejecta and a pre-existing circumstellar wind from the red giant.
\item {\itshape U Sco type (U Sco, V394 CrA, CI Aql and V2487 Oph)\/} have short orbital periods ($\sim$1 d). The secondaries are main-sequence stars or sub-giants. Outburst ejection velocities are extremely high. They decline very rapidly, especially U Sco which is the fastest nova observed. The quiescent spectra are dominated by He lines unlike those in typical novae.
\item {\itshape T Pyx type (T Pyx and IM Nor)\/} have orbital periods $\sim$hours similar to CNe and show relatively slow decay with oscillations in the transition region of the light curves.
\end{enumerate}

RNe have been proposed as one of the primary candidates for the progenitors of Type Ia Supernovae (SNe Ia) through the single degenerate scenario in which the white dwarf can grow in mass up to the Chandrasekhar limit (1.4$M_{\odot}$) if the accreted mass is larger than the ejected mass after eruption [3].

It is thus important to investigate the nature of central binary systems of CNe and RNe to determine the relation between the parameters of the central system  (e.g. type of secondary) and outburst type and ultimately ascertain the population of novae that might be progenitors of Type Ia SNe.

\section{Selection of Initial Targets}
The proposal that RNe occupy a region separated from CNe in an amplitude-$t_{3}$ diagram suggested by ref.[4] was adopted. The low amplitude results from the existence of an evolved secondary and/or high mass transfer rate in the quiescent system. The 93 novae with $V$ amplitudes from ref.[5] and 43 novae with photographic amplitudes from ref.[6] were combined and plotted on an outburst amplitude ($A'$) versus rate of decline ($t_{3}$) diagram as shown in Figures 1 and 2 respectively, from which the initial targets that are suspected to be RNe candidates were selected.

\begin{figure}[pb!]
\begin{minipage}{16pc}
\includegraphics[width=14pc]{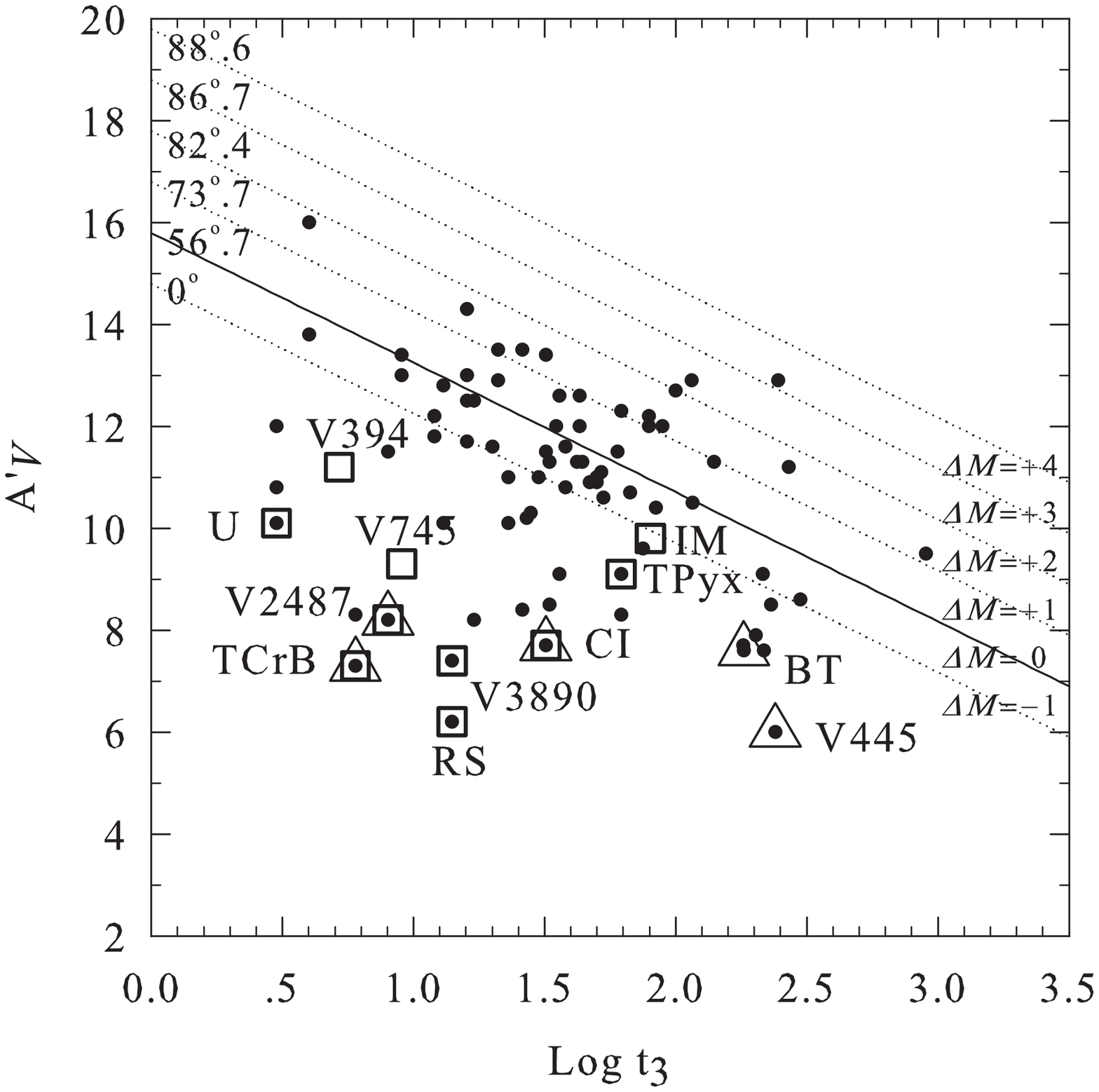}
\caption{\label{label}Target selection from $V$ amplitudes and rate of decline.}
\end{minipage}\hspace{1pc}%
\begin{minipage}{16pc}
\includegraphics[width=14pc]{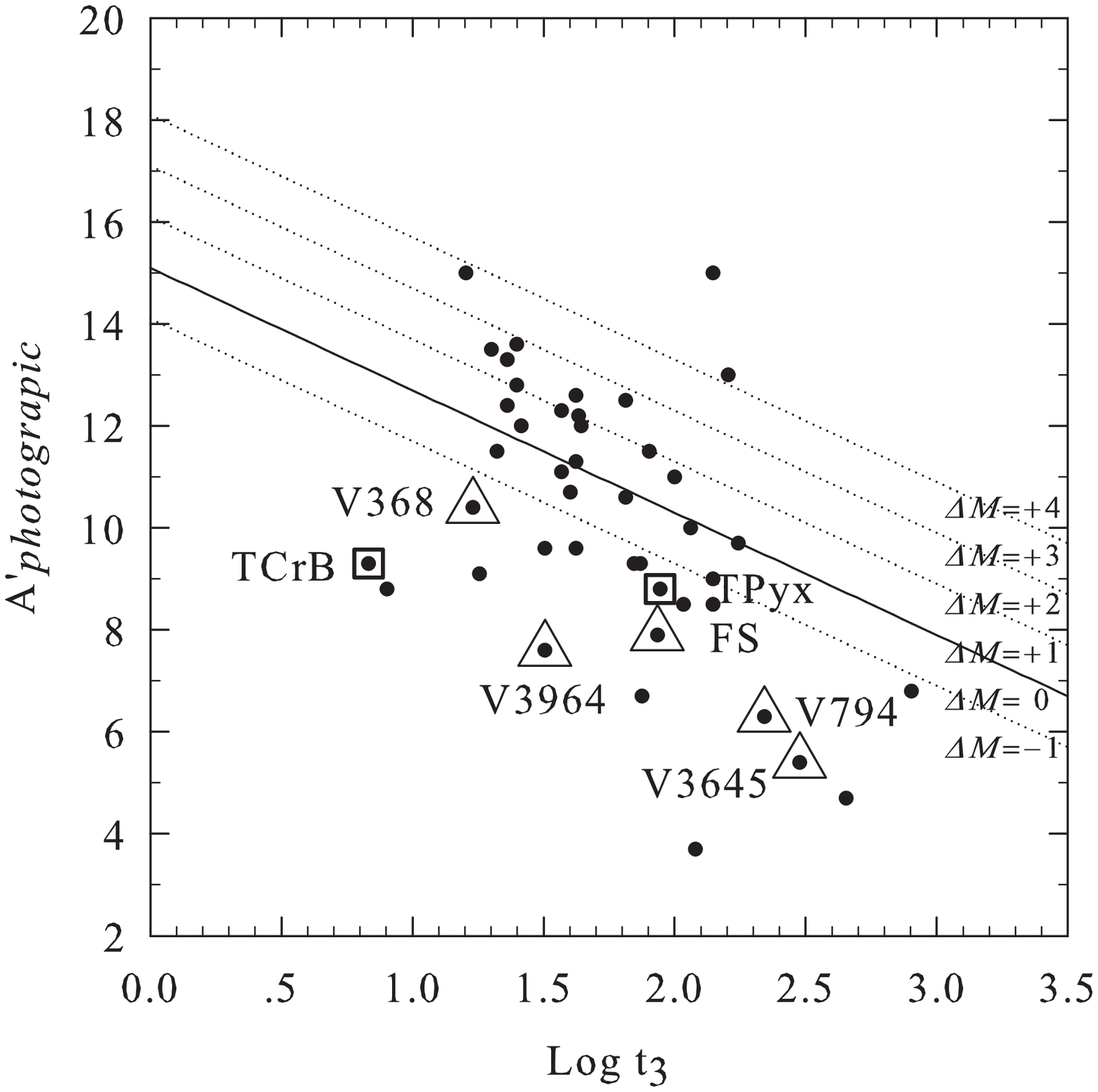}
\caption{\label{label}Target selection from photographic amplitudes and rate of decline.}
\end{minipage}
\end{figure}

Figure 1 shows that all Galactic RNe occupy the bottom area of the plot representing the low $V$ amplitude region among the CNe. A solid diagonal line represents a typical CNe trend line calculated by using the following MMRD relation given in ref.[7]
\[
    A'= [\overline{M}_{V}]_{min} - [M_{V}]_{max} , [M_{V}]_{max}= a\log{t_{3}} + b
    \label{amp}
\]
where $t_{3}$ is a time (in days) that a nova takes to decline 3 magnitudes from peak, $[M_{V}]_{max}$ is a magnitude at maximum brightness, and $[\overline{M}_{V}]_{min}$ = 3.8 is an average $[M_{V}]_{min}$ over all speed class adopted from ref.[8], and coefficients $a$ and $b$ for $V$ magnitude are adopted as 2.54 and 11.99 from ref.[7]. Since the $A'$ values are uncorrected for inclination of the discs, the effects of inclination are shown by the diagonal lines. This variation in absolute magnitude of an accretion disc can be expressed as a function of inclination given by ref.[9] as follows
\[
    \Delta{M_{V}(i)} = -2.5\log{[\cos{i}+\frac{3}{2}\cos^{2}{i}]}
    \label{deltam}
\]
where $\Delta$$M_{V}$=-1, 0, 1, 2, 3 and 4 correspond to inclinations 0$^{\circ}$, 56$^{\circ}$.7, 73$^{\circ}$.7, 82$^{\circ}$.4, 86$^{\circ}$.7, and 88$^{\circ}$.6 respectively.

Figure 2 also shows the same aspect in photographic amplitudes. The typical CNe trend line was plotted from the same equation but with coefficients $a$=2.4 and $b$=11.3 from ref.[7] for photographic magnitude.

Initial targets including known RNe were selected according to their amplitude difference from the typical CNe line from both catalogues. The further the amplitude from the CNe line, the higher priority the object was given. The first 10 targets, which were observable by the Liverpool Telescope (LT) during 2011 May, arranged by this priority are T CrB, V2487 Oph, CI Aql, V3964 Sgr, V3645 Sgr, V445 Pup, V794 Oph, FS Sct, BT Mon and V368 Aql.

\begin{figure}[h!]
\begin{minipage}{16pc}
\includegraphics[width=14pc]{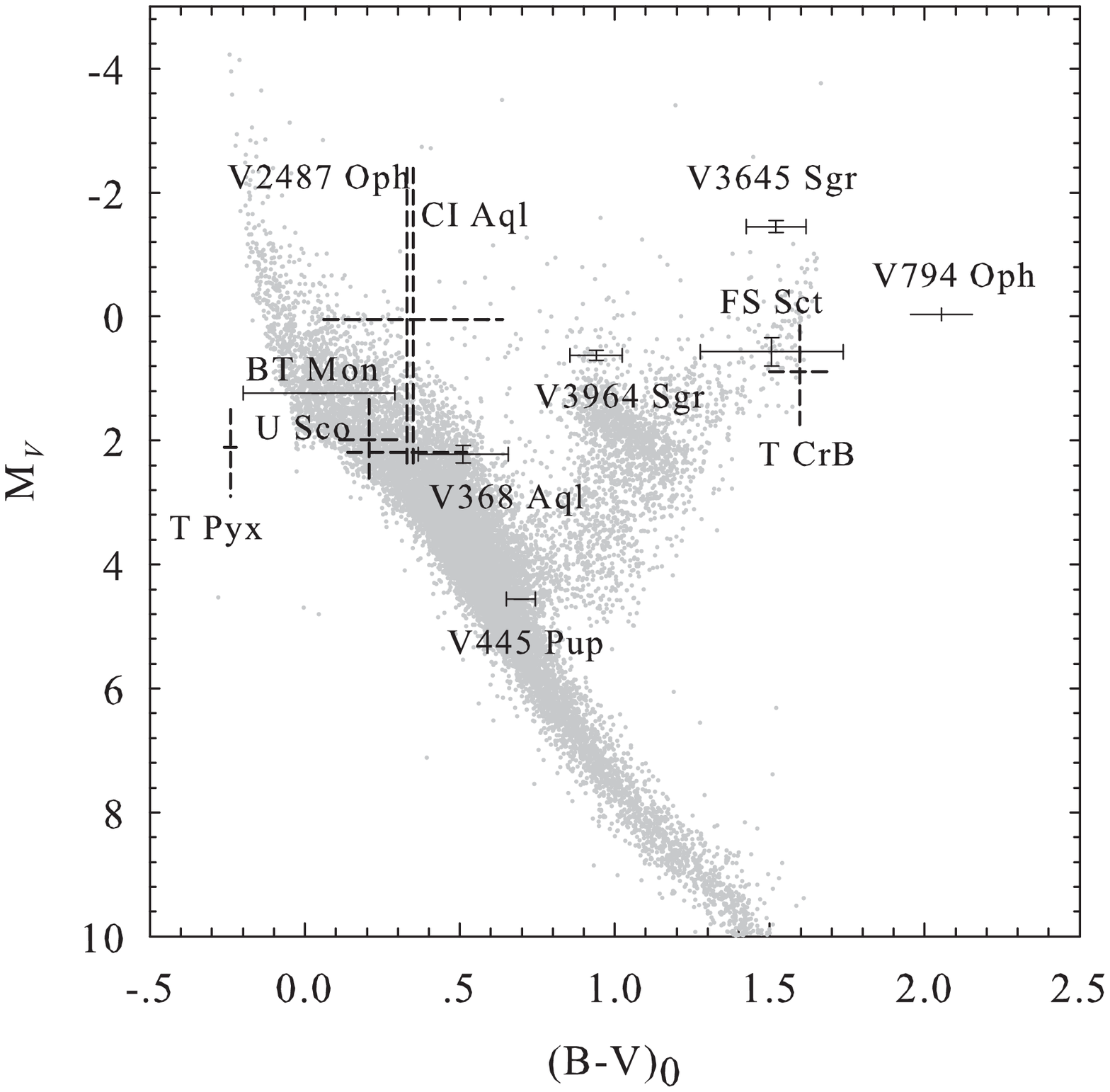}
\caption{\label{label}CMD showing 10 targets compared with stars plotted from $Hipparcos$ data. Data for T Pyx and U Sco are adopted from ref.[10]. }
\end{minipage}\hspace{1pc}%
\begin{minipage}{16pc}
\includegraphics[width=14pc]{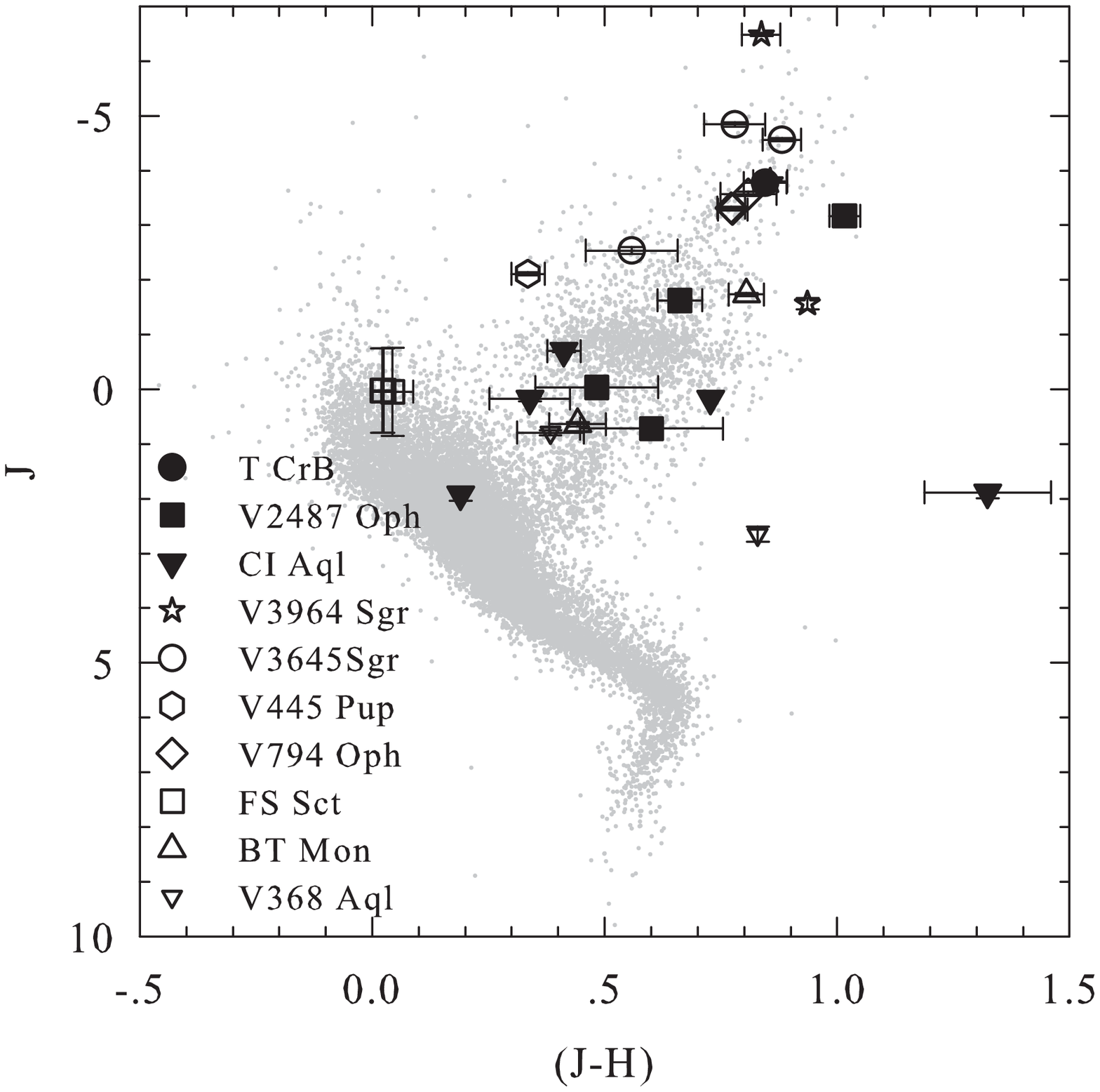}
\caption{\label{label}CMD showing 10 targets compared with stars generated by cross-correlating the $Hipparcos$ and 2MASS catalogues.}
\end{minipage}
\end{figure}

\section{Observations and Results}
The $B,V,u',r',i',z'$ photometric observations were carried out at the fully robotic 2-metre Liverpool Telescope (LT) in La Palma, Spain during 2011 May. Figure 3 shows a colour-magnitude diagram (CMD) of 10 targets compared to Hipparcos data from ref.[11]. Error bars in $(B-V)_{0}$ for 7 CNe were calculated from the all-sky extinction maps provided by three sources: refs [12], [13], and [14]. Error bars in $(B-V)_{0}$ for RNe were adopted from ref.[10]. Four targets including V3645 Sgr, V794 Oph and FS Sct are found to lie near the red giant branch. Figure 4 shows their positions on a near-infrared CMD from ref.[15]. V3645 Sgr and V794 Oph in particular appear to contain red giant secondaries. Comparing positions on CMDs of our targets with those presented in ref.[16] V3645 Sgr and V794 Oph may be representatives of the RS Oph type RNe. Meanwhile BT Mon and V368 Aql may be representatives of the short period U Sco type.

Follow-up spectroscopic observations have now commenced using FRODOSpec (Fibre-fed RObotic Dual-beam Optical Spectrograph) on LT. Figure 5 shows a preliminary relative flux-calibrated spectrum of T CrB showing the emission is dominated by the red giant secondary star. Similarly, Figure 6 shows a preliminary spectrum of V2645 Sgr, again dominated by a cool secondary star.

\begin{figure}[h!]
\begin{minipage}{16pc}
\includegraphics[width=15pc]{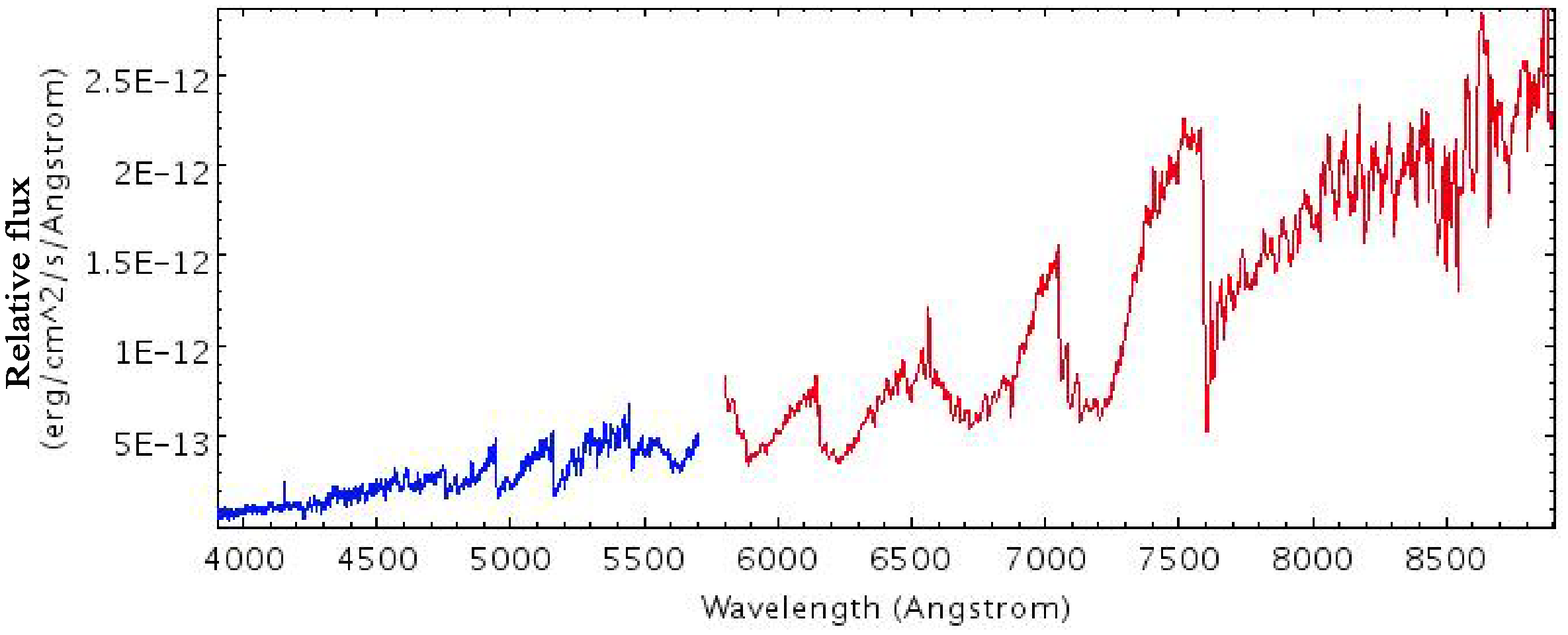}
\caption{\label{label}Spectrum of T CrB from FRODOSpec on the LT. Spectra is not photometric calibrated and atmospheric absorptions are not removed.}
\end{minipage}\hspace{1pc}%
\begin{minipage}{16pc}
\includegraphics[width=15pc]{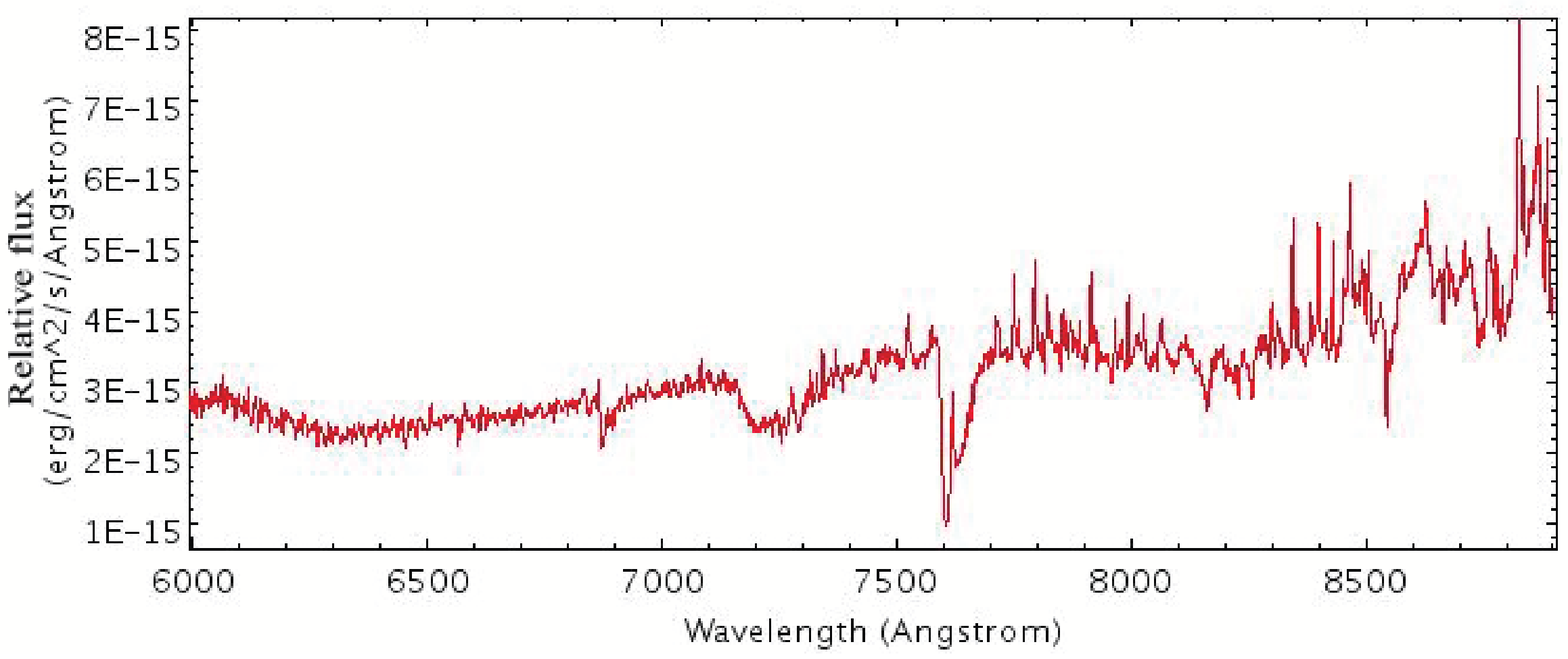}
\caption{\label{label}Spectrum of V3645 Sgr in the red arm of FRODOSpec. Spectra is not photometric calibrated and atmospheric absorptions are not removed.}
\end{minipage}
\end{figure}

\section{Conclusion}
We selected 10 suspected RNe based on their low outburst amplitudes and and accessibility for observation have commenced a programme of photometric and spectroscopic observations. Positions of targets on CMDs reveal that V3645 Sgr and V794 Oph may join the RS Oph type RNe; while BT Mon and V368 Aql may join the short period U Sco type. Five candidates now have LT spectra. Initial future work is to ascertain the spectral type of the companion stars by investigating characteristics of the spectra and also to compare to those of the known Galactic RNe.

\end{document}